\documentclass[lettersize,journal]{IEEEtran}
\usepackage{amsmath,amsfonts}
\usepackage{algorithm}
\usepackage{algorithmic}
\usepackage{array}
\usepackage[caption=false,font=normalsize,labelfont=sf,textfont=sf]{subfig}
\usepackage{textcomp}
\usepackage{stfloats}
\usepackage{url}
\usepackage{multirow}
\usepackage{tabularx,booktabs} %Self added 
\usepackage{verbatim}
\usepackage{graphicx}
\usepackage{cite}
\usepackage{makecell}
\usepackage{url}
\usepackage{relsize} %Self added
\usepackage{color}
\usepackage{graphicx}
\makeatletter
\let\NAT@parse\undefined
\makeatother
\usepackage{hyperref}  %hyperref still needs to be put at the end!

\definecolor{b}{rgb}{0,0,0} %%Pure Version

\begin{document}

\title{\LARGE Deep Reinforcement Learning-Based Cooperative Rate Splitting for Satellite-to-Underground Communication Networks}

\author{Kaiqiang~Lin,~\IEEEmembership{Member,~IEEE,}
        Kangchun Zhao, and 
        Yijie~Mao,~\IEEEmembership{Member,~IEEE}
% \vspace{-10mm}    
\thanks{This work was supported in part by the National Nature Science Foundation of China under Grant 62571331. (\textit{Corresponding author: Yijie Mao})}  

\thanks{K. Lin is with the Division of Computer, Electrical and Mathematical Sciences and Engineering, King Abdullah University of Science and Technology, Saudi Arabia (E-mail: kaiqiang.lin@kaust.edu.sa).}

\thanks{K. Zhao and Y. Mao are with the School of Information Science and Technology, ShanghaiTech University, Shanghai, China (E-mail: zhaokch12022@shanghaitech.edu.cn, maoyj@shanghaitech.edu.cn).}
}


\maketitle

\begin{abstract}
Reliable downlink communication in satellite-to-underground networks remains challenging due to severe signal attenuation caused by underground soil and refraction in the air-soil interface. To address this, we propose a novel cooperative rate-splitting (CRS)-aided transmission framework, where an aboveground relay decodes and forwards the common stream to underground devices (UDs). Based on this framework, we formulate a max-min fairness optimization problem that jointly optimizes power allocation, message splitting, and time slot scheduling to maximize the minimum achievable rate across UDs. To solve this high-dimensional non-convex problem under uncertain channels, we develop a deep reinforcement learning solution framework based on the proximal policy optimization (PPO) algorithm that integrates distribution-aware action modeling and a multi-branch actor network. \textcolor{b}{Simulation results under a realistic underground pipeline monitoring scenario demonstrate that the proposed approach achieves a higher average max-min rate than conventional benchmark strategies across varying numbers of UDs and diverse underground conditions.}

%Satellite-to-underground networks offer tremendous economic and societal benefits in remote agriculture, industrial monitoring, and post-disaster rescue. However, ensuring reliable downlink communication in such networks remains challenging due to severe signal attenuation caused by underground soil and refraction in the air-soil interface. To address this, we propose a novel cooperative rate-splitting (CRS)-aided satellite-to-underground transmission framework, where an aboveground relay decodes and forwards the common stream to underground devices (UDs). Based on this framework, we formulate a max-min fairness optimization problem that jointly optimizes power allocation, message splitting, and time slot scheduling to maximize the minimum achievable rate across UDs. To handle channel uncertainty, we develop a deep reinforcement learning solution framework based on the proximal policy optimization (PPO) algorithm to address the formulated problem. Simulation results in a realistic underground pipeline monitoring scenario demonstrate that the proposed PPO-based CRS approach achieves superior max-min rate performance compared to conventional benchmark strategies.
\end{abstract}


\section{Introduction}
\IEEEPARstart{S}{atellite}-to-underground networks have been recognized as a promising communication paradigm that enables direct or relayed data transmission between satellites and devices located below ground level. It facilitates subterranean monitoring in hard-to-reach or disaster-stricken areas, supporting applications such as remote smart agriculture, underground pipeline monitoring, and post-disaster rescue~\cite{LinMag}. Although previous studies~\cite{LinMag, LinWCLDRL, LinAdhoc} have demonstrated the feasibility of uplink communication from underground devices (UDs) to low-Earth-orbit (LEO) satellites, the realization of reliable downlink communication in satellite-to-underground networks remains largely unexplored. 

In the meanwhile, rate-splitting multiple access (RSMA), which employs linearly precoded rate-splitting at the transmitter to divide user messages into common and private parts, and applies successive interference cancellation (SIC) at the receivers to sequentially decode common and private streams, has emerged as a more efficient and robust downlink interference management strategy than space division multiple access (SDMA) and power-domain non-orthogonal multiple access (NOMA)~\cite{MaoRSMAsurvey}. Moreover, RSMA is technically feasible for UDs due to its single-layer or even SIC-free decoding architecture and its compatibility with wireless energy transfer technologies for sustainable operation. Therefore, RSMA is a promising solution for enabling downlink communications in satellite-to-underground networks, offering potential advantages in spectral and  energy efficiency enhancement. However, such application remains unexplored in prior work. One fundamental characteristic of RSMA is that the common stream must be decoded by multiple users, which constrains the achievable rate to that of the the worst-case user. This limitation is more pronounced in satellite-to-underground networks due to the severe attenuation from LEO satellites to UDs caused by the severe signal absorption in soil and refraction loss in the air-soil interface. 

To address these research challenges, in this work, we extend the cooperative rate-splitting (CRS) strategy proposed in~\cite{CRSTWC} to the satellite-to-underground networks, where an aboveground relay (AR) with better channel conditions forwards the decoded common stream from the LEO satellite to the weaker UDs, thereby enhancing the UDs’ ability to decode the common stream under harsh underground environments. Based on the proposed model, we investigate the joint optimization of power allocation, message splitting, and time-slot scheduling to maximize the minimum achievable rate among UDs. Existing studies in CRS typically assume perfect channel state information (CSI) or CSI distribution at the transmitter, this assumption, however, becomes impractical in our scenario due to three key challenges: (1) the highly dynamic nature of underground channels caused by time-varying soil properties, (2) significant propagation delays inherent in satellite links, and (3) fast-fading conditions in the air-soil interface. 

These unique characteristics necessitate a novel resource allocation framework that can operate effectively under uncertain channel conditions. Recently, deep reinforcement learning (DRL) has emerged as a powerful paradigm for intelligent decision-making in RSMA~\cite{PPOSDWCL, GreedyTcom} and RSMA-based satellite–terrestrial networks~\cite{PPOLEOWCL}, without requiring prior channel information. \textcolor{b}{Nevertheless, existing DRL-based RSMA studies do not consider CRS, whereas prior CRS works typically rely on conventional optimization frameworks under ideal CSI assumptions~\cite{CRSTcom, CRS2025}. Motivated by this gap, we employ a highly effective DRL algorithm, namely proximal policy optimization (PPO), to enable coordinated decision-making over power control, message splitting, and time-slot allocation. This joint design balances resource efficiency and fairness while maximizing the worst-case rate among UDs under dynamic and uncertain channel conditions. To the best of our knowledge, this is the first work to investigate the effectiveness of DRL for CRS-aided satellite-to-underground downlink communications.} Through extensive simulation results, we reveal the superiority of our proposed PPO-based CRS approach over three well-established benchmarks in realistic underground pipeline monitoring scenarios.

\section{System Model and Problem Formulation}
\begin{figure}[!t]
\centering
\includegraphics[width=3.45in]{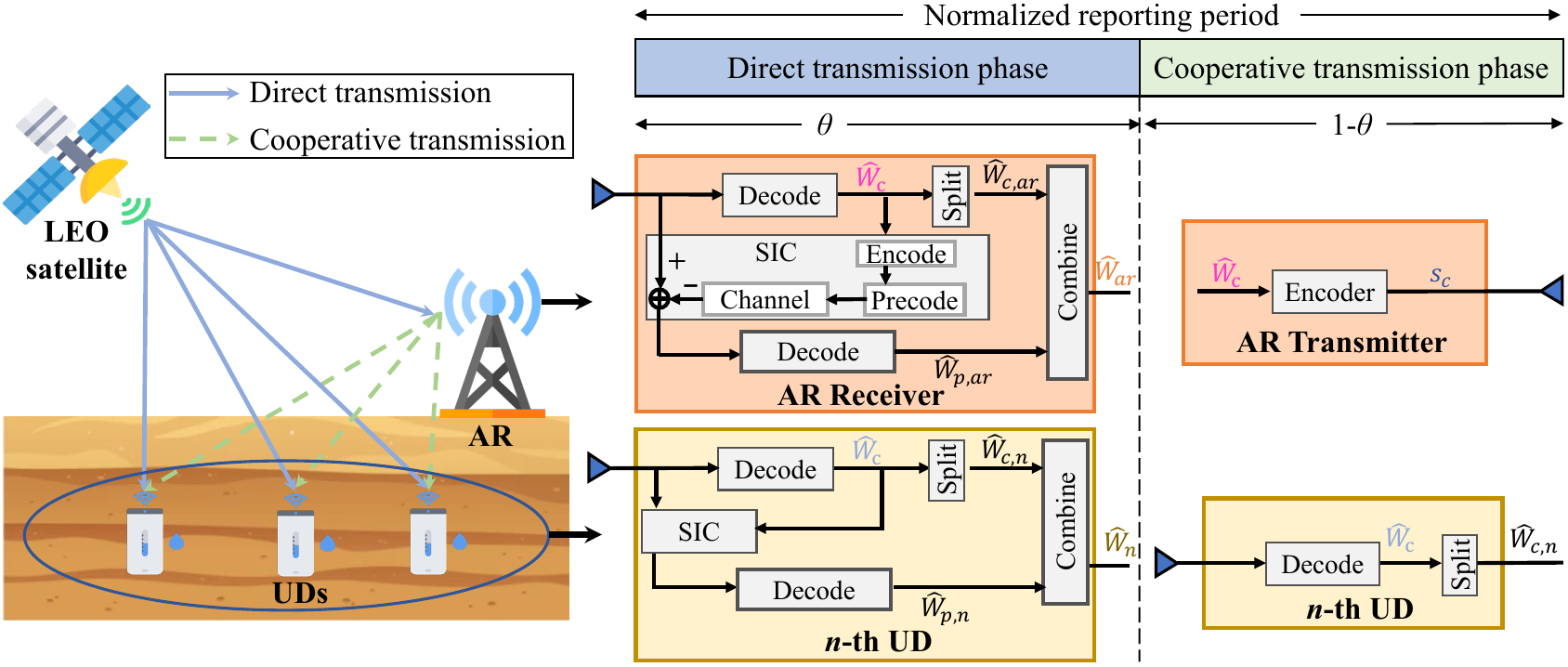}
\caption{The proposed CRS-aided satellite-to-underground network system with the corresponding time slot allocation for its two transmission phases.}
\label{fig_1}
\end{figure}

Consider a satellite-to-underground downlink communication system as depicted in Fig.~\ref{fig_1}, where a $Q$-antenna LEO satellite serves a single-antenna AR and $N$ single-antenna UDs, indexed by $\mathcal{N}=\{1, 2, \ldots, N\}$, all buried at the same depth $d_u$. The CRS transmission scheme is enabled to enhance downlink communication. Specifically, in each normalized coherent transmission period, the LEO satellite first transmits signals to both the AR and the UDs during the direct (or first) transmission phase. Subsequently, the AR employs the non-regenerative decode-and-forward protocol to forward the received signals to the UDs during the cooperative (or second) transmission phase. A fraction of time $\theta$ is allocated to direct transmission phase, while the remaining portion $1 - \theta$ is allocated to cooperative transmission phase.

We assume that the LEO satellite holds a total of $N+1$ messages, denoted by ${W_{ar}, W_{1}, \ldots, W_{N}}$, intended for the AR and the $N$ UDs, respectively. In accordance with the 1-layer RSMA principle, each message is divided into a common part and a private part. The common parts ${W_{c, ar}, W_{c, 1}, \ldots, W_{c, N}}$ are jointly encoded into a single common stream $s_c$ using a common codebook, which is intended to be decoded by the AR and all UDs. The private parts ${W_{p, ar}, W_{p, 1}, \ldots, W_{p, N}}$ are independently encoded into private streams ${s_{ar}, s_{1}, \ldots, s_{N}}$, each targeting a specific receiver. Assuming a linear precoding scheme, the transmit signal at the LEO satellite (in the first transmission phase) is given by 
\begin{equation}
    \mathbf{x} = \sqrt{P_c} \mathbf{w}_c s_c + \sqrt{P_{ar}} \mathbf{w}_{ar} s_{ar}+ \sum\nolimits_{n=1}^{N} \sqrt{P_{n}}  \mathbf{w}_{n} s_{n},
\end{equation}
where $P_c$, $P_{ar}$, and $P_n$ are the transmit power allocated to the common stream, the private stream for the AR, and the private stream for the $n$-th UD, respectively. $\mathbf{w}_c$, $\mathbf{w}_{ar}$, and $\mathbf{w}_n$ are the corresponding precoding vectors. Accordingly, the signals received by the AR and the $n$-th UD during the first transmission phase are expressed as
\begin{align}
    y_{ar} &= \mathbf{h}_{ar}^{H} \mathbf{x} + n_{ar}, \\
    y_n &= \mathbf{h}_n^{H} \mathbf{x} + n_n,
\end{align}
where $n_{ar} \sim \mathcal{CN}(0, \sigma_{ar}^2)$ and $n_n \sim \mathcal{CN}(0, \sigma_n^2)$ denote the additive white Gaussian noise (AWGN) at the AR and the $n$-th UD, respectively. $\mathbf{h}_{ar} \in \mathbb{C}^{Q \times 1}$ and $\mathbf{h}_{n} \in \mathbb{C}^{Q \times 1}$ represent the channels from the LEO satellite to the AR and to the $n$-th UD, respectively. They are modeled as~\cite{LinMag}
\begin{align}
    \mathbf{h}_{ar} &= \boldsymbol{\delta}_{ar} \sqrt{G_s G_{ar}\left(\frac{c}{4 \pi f d_{s2a}}\right)^2}, \\
    \mathbf{h}_n &=  \boldsymbol{\delta}_n \sqrt{\frac{G_s G_n}{L_{n}^{r}L_{n}^{u}}\left(\frac{c}{4 \pi f d_n^{s2g}}\right)^2},
\end{align}
where $\boldsymbol{\delta}_{ar}$ and $\boldsymbol{\delta}_n$ denote the small-scale fading channel vectors from the LEO satellite to the AR and the $n$-th UD, respectively, each following a Rician distribution. $G_s$, $G_{ar}$, and $G_n$ are the antenna gains of the LEO satellite, the AR, and the $n$-th UD, respectively. $c$ is the speed of light, $f$ is the carrier frequency, $d_{s2a}$ and $d_{n}^{s2u}$ denote the air propagation paths from the LEO satellite to the AR and to the $n$-th UD, respectively. $L_{n}^{r}$ and $L_{n}^{u}$ represent the refraction loss at the air–soil interface and the attenuation in underground soil, respectively. According to the validated channel model developed in~\cite{Undergroundfield, LinBSWPUSNs}, they are expressed as
\begin{align}
\label{Rpathloss} L_{n}^{r}&=\left(\left(\sqrt{\left(\sqrt{\varepsilon'^{2}+\varepsilon''^{2}}+\varepsilon' \right) / 2}+1\right)/4\right)^2, \\
\label{soilpathloss} L_{n}^{u}&=\left(2 \beta d_n^{soil}/e^{-\alpha d_n^{soil}}\right)^{2}.
\end{align}
Herein, $d_{n}^{soil}$ is the underground soil propagation distance from ground surface to $n$-th UD. Since the permittivity of soil is much larger than air, most RF signal energy from the above-ground sink will be reflected back if the incident angle is large. Therefore, we only consider the signal with a small incident angle, and the refracted angle is close to zero during the signal propagation from air to underground soil. Thus, in this study, we assume that the propagation in the soil is vertical, i.e., $d_{n}^{soil}=d_{u}$. Additionally, $\alpha$ and $\beta$ respectively represent the attenuation and phase shifting constants, which are given as
\begin{align}
\alpha &= 2 \pi f \sqrt{\frac{\mu_{r} \mu_{0} \varepsilon' \varepsilon_{0}} {2}\left[\sqrt{1+\left(\varepsilon''/\varepsilon'\right)^{2}}-1 \right]}, \\
\label{beta}\beta &= 2 \pi f \sqrt{\frac{\mu_{r} \mu_{0} \varepsilon' \varepsilon_{0}} {2}\left[\sqrt{1+\left(\varepsilon''/\varepsilon'\right)^{2}}+1 \right]}.
\end{align}
Herein, $\mu_{r}$ is the soil’s relative permeability, $\mu_{0}$ is the free-space permeability, $\varepsilon_{0}$ is the free space permittivity, and $\varepsilon'$ and $\varepsilon''$ are the real and imaginary parts of the soil’s relative permittivity, respectively, i.e., $\varepsilon = \varepsilon' + j\varepsilon''$. Note that $\varepsilon$ can be calculated by the accurate mineralogy-based soil dielectric model~\cite{MBSDM}. For this, only three input parameters are required: the volumetric water content (VWC), the operating frequency of the RF signals, and the percentage of clay in soil.

In the direct transmission phase, the common stream $s_c$ is decoded firstly while treating the private streams as noise. Thus, the instantaneous signal to interference plus noise ratios (SINRs) of decoding $s_c$ at the AR and the $n$-th UD are respectively given by
\begin{align}
    \gamma_{c,ar}^{D} &=\frac{P_c \left|\mathbf{h}_{ar}^{H} \mathbf{w}_c\right|^2}{ P_{ar}\left|\mathbf{h}_{ar}^{H} \mathbf{w}_{ar}\right|^2 + \sum_{i=1}^{N} P_i \left|\mathbf{h}_{ar}^H \mathbf{w}_i\right|^2 +\sigma_{ar}^2}, \\
    \gamma_{c,n}^{D} &= \frac{P_c \left|\mathbf{h}_n^H \mathbf{w}_c\right|^2}{P_{ar}\left|\mathbf{h}_{n}^{H} \mathbf{w}_{ar}\right|^2+\sum_{i=1}^{N} P_i \left|\mathbf{h}_n^H \mathbf{w}_i\right|^2 +\sigma_{n}^2}.
\end{align}
Herein, $P_c \left|\mathbf{h}_{ar}^{H} \mathbf{w}_c\right|^2$ and $P_c \left|\mathbf{h}_n^H \mathbf{w}_c\right|^2$ represent for the target received power of the common stream for the AR and the $n$-th UD, respectively, $P_{ar}\left|\mathbf{h}_{ar}^{H} \mathbf{w}_{ar}\right|^2$ and $P_{ar}\left|\mathbf{h}_{n}^{H} \mathbf{w}_{ar}\right|^2$ denote the interference caused by the private stream intended for the AR, $\sum_{i=1}^{N} P_i \left|\mathbf{h}_{ar}^H \mathbf{w}_i\right|^2$ and $\sum_{i=1}^{N} P_i \left|\mathbf{h}_n^H \mathbf{w}_i\right|^2$ account for the interference from the private streams transmitted to all UDs, which are treated as noise when decoding the common stream, while $\sigma_{ar}^2$ and $\sigma_{n}^2$ denote the AWGN power at the AR and the $n$-th UD, respectively.

Accordingly, the achievable rates of the common stream in the direct transmission phase at the AR and $U_n$ are $R_{c,ar}^{D} = \theta \log_2(1+\gamma_{c,ar}^{D})$ and $R_{c,n}^{D} = \theta \log_2 (1+\gamma_{c,n}^{D})$, respectively.

After performing the SIC and removing the common stream from the received signal, the SINRs of decoding private stream at the AR and the $n$-th UD in the direct transmission phase are respectively given by
\begin{align}
    \gamma_{p,{ar}}^{D} &= \frac{P_{ar} \left|\mathbf{h}_{ar}^H \mathbf{w}_{ar}\right|^2}{\sum_{i=1}^{N} P_i \left|\mathbf{h}_{ar}^H \mathbf{w}_i\right|^2 + \sigma_{ar}^2},\\
    \gamma_{p,n}^{D} &= \frac{P_n \left|\mathbf{h}_n^H \mathbf{w}_n\right|^2}{P_{ar}\left|\mathbf{h}_{n}^{H} \mathbf{w}_{ar}\right|^2 + \sum_{i=1, i \neq n}^{N} P_i |\mathbf{h}_n^H \mathbf{w}_i|^2 + \sigma_{n}^2},
\end{align}
where $P_{ar} \left|\mathbf{h}_{ar}^H \mathbf{w}_{ar}\right|^2$ and $P_n \left|\mathbf{h}_n^H \mathbf{w}_n\right|^2$ represent the desired received power of the private stream for the AR and the $n$-th UD, respectively, while $\sum_{i=1, i \neq n}^{N} P_i |\mathbf{h}_n^H \mathbf{w}_i|^2$ accounts for the interference from other UDs’ private streams, explicitly excluding the $n$-th UD’s own private stream. The corresponding achievable rate of the private stream in the direct transmission phase at the AR and $U_n$ are $R_{p,ar}^{D} = \theta \log_2 (1+\gamma_{p,ar}^{D})$ and $R_{p,n}^{D} = \theta \log_2 (1+\gamma_{p,n}^{D})$, respectively.

In the cooperative transmission phase, the AR re-encodes its decoded $s_c$ by employing a different codebook from that of the LEO satellite, and then retransmits it to all UDs through a transmit power $P_R$. Note that the LEO satellite and all UDs remain silent. Since the transmission in this phase proceeds through a single-input single-output channel, the achievable rate of decoding the common stream at the $n$-th UD is 
\begin{equation}
    R_{c, n}^{C} = \operatorname{min} \left(\left\{R_{c,ar}^{D}\right\}, \left \{(1-\theta)\log_2 \left(1+\frac{P_{R} |h_{{ar}, n}|^2}{\sigma_{n}^2}\right)\right\} \right),
\end{equation}
where ${h}_{{ar}, n}$ is the channel gain from the AR to the $n$-th UD. It is given by~\cite{Undergroundfield, LinBSWPUSNs}
\begin{equation}
    h_{{ar}, n} = \delta_{ar, n} \sqrt{\frac{G_{ar} G_u}{L_{n}^{r}L_{n}^{u}} \left(\frac{c}{4 \pi f d_{n}^{a2u}}\right)^2},
\end{equation}
where $\delta_{ar, n}$ denotes the small-scale fading from the AR to the $n$-th UD, modeled by a Rician distribution, while $d_{n}^{a2g}$ denotes the air propagation distance from the AR to the $n$-th UD.

After the cooperative transmission phase, all UDs combine the decoded common stream decoded in both phases. To ensure that both the AR and all UDs can successfully decode $s_c$, the achievable rate of the common stream is given by
\begin{align}
    R_{c} &= \operatorname{min} \left(\left\{R_{c,ar}^{D}\right\}, \left\{R_{c,n}^{D} + R_{c, n}^{C}| n \in \mathcal{N}\right
    \}\right).
\end{align}
As $R_c$ is shared by the AR and all UDs for the transmission of common stream $s_c$, we have $C_{ar}+\sum_{n=1}^{N} C_{n}=R_c$, where $C_{ar}$ and $C_n$ represent the portions of $R_c$ allocated for transmitting $W_{c, ar}$ and $W_{c, n}$, respectively. After decoding and removing $s_c$ from the received signal, the AR and the $n$-th UD proceed to decode their respective private streams. Therefore, the total achievable rates of the AR and the $n$-th UD are expressed as $R_{ar}^{\text{tot}} = R_{p,{ar}}^{D}+C_{ar}$ and $R_{n}^{\text{tot}} = R_{p,n}^{D}+C_n$.

% \subsection{Max-Min Fairness Problem Formulation}
By enabling the AR forward its decoded common message to the UDs, the proposed CRS mechanism enhances the achievable rate of the common stream, effectively mitigating the severe attenuation challenges in satellite-to-underground downlink communication compared with SDMA and RSMA schemes without an AR. In this framework, we further emphasize the user fairness by maximizing the worst available rate among all UDs. This is achieved by jointly optimizing the transmit power vector allocated to the common and private streams $\mathbf{p} = \{P_c, P_{ar}, P_1, \ldots, P_N\}$, the common rate vector as $\mathbf{c} = \{C_{ar}, C_1, C_2, \ldots, C_N\}$, and the time slot allocation $\theta$. To focus on optimizing these resource allocations, a fixed precoding scheme is adopted: the common stream is precoded using a maximum ratio transmission vector, while private streams employ normalized matched filtering. Accordingly, the max-min rate optimization problem for the CRS-aided satellite-to-underground downlink system is formulated as
\begin{subequations}\label{p1}
\begin{align} 
({\rm{P1}}):~{\mathop {\max} \limits_{\mathbf{p}, \mathbf{c}, \theta}~\min_{n \in \mathcal{N}}} ~&{R_{n}^{\text{tot}}}\\  
\text{s.t.}~ \label{c1_1} &C_{ar}+\sum\nolimits_{n=1}^{N} C_n \leq R_c,\\
             \label{c1_2} &P_c+P_{ar}+\sum\nolimits_{n=1}^{N} P_n \leq P_t,\\
             \label{c1_3} &0 \leq \theta \leq 1,\\
             \label{c1_4} & \mathbf{c} \geq 0, 
\end{align}
\end{subequations}
where constraint~\eqref{c1_1} guarantees that the common stream is successfully decoded by the AR and all UDs, constraint~\eqref{c1_2} is the transmit power constraint at the LEO satellite, constraint~\eqref{c1_3} impose $\theta$ ranges from $[0, 1]$, and constraint~\eqref{c1_4} is guarantee the non-negative rate of the common stream. 

The optimization problem (P1) is non-convex and it is infeasible to find the optimal solution within the exhaustive searching method due to the continuous value of power control, common rate split, and time slot allocation. Furthermore, the LEO satellite lacks knowledge of the channel state distribution because of the dynamic and uncertain characteristics of the satellite-to-underground communication environment, making it difficult to apply conventional optimization methods effectively. To address these challenges, we adopt a DRL-based approach in the next section to find the optimal solution for the problem (P1) and enable adaptive resource allocation without requiring prior knowledge of the environment.

\section{DRL Framework for Resource Allocation}
In this section, we design a DRL-based optimization framework to jointly optimize the power control vector $\mathbf{p}$, the common rate split $\mathbf{c}$, and the time slot allocation $\theta$, with the objective of maximizing the minimum data rate among UDs in the CRS-assisted satellite-to-underground system. The essence of DRL lies in trial-and-error interactions between an agent and a dynamic environment. Concretely, we consider the LEO satellite as the agent, which observes the environment state $s_t$ and selects an action $a_t$ according to a policy $\pi$ at each time step $t$. For our design, each interaction between the agent and the environment occurs per reporting period, with each reporting period corresponding to a single time step. Upon executing action $a_t$, the agent receives a reward $r_{t}$ reflecting the quality of its decision, and the environment transits to the next state $s_{t+1}$.

% \subsection{Key Components of DRL}
The three key elements involved in the DRL interaction process are defined as follows.
\begin{enumerate}
    \item Action: At time step $t$, the action executed by the agent is defined as $a_t = [\mathbf{p}, \mathbf{c}, \theta]_t$, where $\mathbf{p}$, $\mathbf{c}$, and $\theta$ denote the transmit power vector, the common rate vector, and the time allocation ratio, respectively. Note that the range of these actions should guarantee the constraints~\eqref{c1_1},~\eqref{c1_2}, ~\eqref{c1_3}, and~\eqref{c1_4}.  
    
    \item State: The state needs to encompass useful information that enables the agent to learn effectively and make appropriate decisions. Here, we define the state to include the decoding rate of the common stream $R_c$, the total achievable rates of the AR and all UDs denoted as $\mathbf{R^\text{tot}} = [R_{ar}^{\text{tot}}, R_{1}^{\text{tot}}, \ldots, R_{N}^{\text{tot}}]$, and the SINR feedback of both the common and private messages from the AR and the UDs represented as $\boldsymbol{\gamma} = [\gamma_{c, ar}^{D}, \gamma_{p, ar}^{D}, \gamma_{c, 1}^{D}, \ldots, \gamma_{c, N}^{D}, \gamma_{p, 1}^{D}, \ldots, \gamma_{p, N}^{D}]$. Accordingly, the state observed by the agent at time step $t$ is given by $s_t = [R_c, \mathbf{R^\text{tot}}, \boldsymbol{\gamma}]_{t-1}$.
    
    \item Rward: Given the formulated optimization problem (P1), the reward function is designed to maximize the minimum rate among all UDs. Therefore, the immediate reward is defined as $r_{t} = \operatorname{min}_{n\in \mathcal{N}} R_{n}^{\text{tot}}$
\end{enumerate}

Since the policy for continuous action spaces cannot be derived using conventional action-value methods (e.g., Q-learning and deep Q-network), we employ the PPO algorithm to determine the optimal resource allocation strategy for the dynamic satellite–to-underground network system. In contrast to value-based approaches, PPO directly optimizes a stochastic policy by updating a neural network $\pi_{\omega}(a|s)$, which models the probability distribution over actions conditioned on the observed state~\cite{PPOAlgorithm}. The PPO framework involves three neural networks: the current actor network $\pi_{\omega}$ with parameters $\omega$, the old actor network $\pi_{\omega_{\text{old}}}$ with parameters $\omega_{\text{old}}$, and the critic network $V_{\phi}$ with parameters $\phi$. The new actor network is responsible for interacting with the environment and generating updated action policies. The old actor network, structurally identical to the new one, retains the previous policy and acts as a baseline to constrain policy updates, thereby ensuring training stability through clipped surrogate objectives. The critic network estimates the state-value function and is used to evaluate the policy generated by the actor networks. %Compared to existing PPO-based RSMA optimization approaches~\cite{PPOLEOTcom}, the proposed PPO framework enables stable and constraint-compliant optimization by incorporating distribution-aware action modeling and a specialized multi-branch actor network designed for time, power, and rate allocation.

The agent (i.e., the LEO satellite) first interacts with the environment using the current actor policy $\pi_{\omega}(a|s)$ for a fixed number of time steps and collects a batch of experience data in the form of $\{(s_t, a_t, r_t, s_{t+1})\}$. Based on these samples, the actor and critic networks are updated multiple times. After the update, the parameters of the old actor network $\omega_{\text{old}}$ are synchronized with the updated parameters $\omega$. Specifically, at each time step $t$, the actor network takes the observed state $s_t$ as the input and outputs the probability distribution over action $a_t$ for the current state. The agent executes an action $a_t$ based on this probability distribution and receives a reward along with the next state $s_{t+1}$. After several time steps, the agent collects a batch of experience data and updates the parameters of the new actor and critic networks via gradient ascent and descent, respectively, i.e., $\omega = \omega + \tau \nabla_{\omega} L_{\text{clip}}(\omega)$ and $\phi = \phi - \tau \nabla_{\phi} L(\phi)$, where $\tau$ is the learning rate and $L_{\text{total}}(\omega, \phi) =  \frac{L(\phi)}{2}-L_{\text{clip}}(\omega)$ is the combined loss function for the new actor network. To ensure stable updates of the actor policy, PPO adopts a clipped surrogate objective function defined as~\cite{PPOAlgorithm}:
\begin{equation}
    L_{\text{clip}}(\omega) = \mathbb{E} \left[ \min \left( r_{\omega} \hat{A}_t,\, \operatorname{clip}(r_{\omega}, 1 - \epsilon, 1 + \epsilon)\, \hat{A}_t \right) \right], \label{CliplossEq}
\end{equation}
where $r_{\omega} = \frac{\pi_{\omega}(a_t|s_t)}{\pi_{\omega_{\text{old}}}(a_t|s_t)}$ denotes the probability ratio between the new and old policies, the clip operation $\operatorname{clip}(\cdot)$ restricts the probability ratio to the interval $[1 - \epsilon, 1 + \epsilon]$, preventing large policy updates that could destabilize training, while $\hat{A}_t$ denotes the advantage function computed using generalized advantage estimation (GAE), and is given by
\begin{equation}
\hat{A}_t = \!\sum\nolimits_{l=0}^{T_{b} - t - 1} \!\left(\eta \varsigma\right)^l \left(r_{t+l} + \eta V_{\phi}(s_{t+l+1})\! -\! V_{\phi}(s_{t+l}) \right), \label{AdvEq}
\end{equation}
where $T_{b}$ is the batch size, $\eta \in (0,1)$ is the discount factor, $\varsigma \in [0,1]$ is the GAE smoothing parameter, while $V_{\phi}(\cdot)$ is the value function predicted by the critic network with parameters $\phi$. The loss function for the critic network is a mean squared error defined as
\begin{equation}
    L(\phi) = \mathbb{E}_t \left[ \left( \sum\nolimits_{\upsilon=0}^{\infty} \eta^\upsilon r_{t+\upsilon+1} - V_{\phi}(s_t) \right)^2 \right]. \label{CriticlossEq}
\end{equation}

\begin{algorithm}[!t]
\small
\caption{PPO-Based CRS Approach}
\label{alg_ppo}
\begin{algorithmic}[1]
\STATE Initialize parameters $\omega$, $\phi$ and set $\omega_{\mathrm{old}} \leftarrow \omega$, experience buffer $\mathcal{D} \leftarrow \emptyset$

\STATE Set hyperparameters: total epochs $T_e=2000$, batch size $T_b=512$, update rounds per episode $M=3$, discount factor $\eta=0.9$, GAE parameter $\varsigma=0.95$, clipping value $\epsilon=0.2$
\STATE Initialize environment and get initial state $s_1$

\FOR{episode = $1$ to $T_{e}$} 
    \FOR{step $t = 1$ to $T_b$}
        \STATE New policy $\pi_\omega$ interacts with the environments and stores $(s_t, a_t, r_t, s_{t+1}, \log \pi_\omega(a_t|s_t))$ in $\mathcal{D}$
        \STATE $s_t \leftarrow s_{t+1}$
    \ENDFOR
    \STATE Compute advantage estimates $\hat{A}_t$ using Eq.~\eqref{AdvEq} based on collected data in $\mathcal{D}$
    \FOR{$m=1$ to $M$}
        \STATE Compute the combined loss $L_{\text{total}}(\omega, \phi) = \frac{L(\phi)}{2} -L_{\text{clip}}(\omega)$ by Eqs.~\eqref{CliplossEq} and~\eqref{CriticlossEq}
        \STATE Perform gradient ascent (for actor) and descent (for critic) updates on $\omega$ and $\phi$ with respect to $L_{\text{total}}(\omega, \phi)$
    \ENDFOR
    \STATE Update old policy parameters: $\omega_{\mathrm{old}} \leftarrow \omega$
    \STATE Clear experience buffer: $\mathcal{D} \leftarrow \emptyset$
\ENDFOR
\end{algorithmic}
\end{algorithm}

\textbf{Algorithm~\ref{alg_ppo}} illustrates the workflow of the proposed PPO algorithm. Compared to existing PPO-based RSMA optimization approaches~\cite{PPOSDWCL, GreedyTcom, PPOLEOWCL}, the proposed PPO framework enables stable and constraint-compliant optimization by leveraging distribution-aware action modeling and a specialized multi-branch actor designed for time, power, and rate allocation.

\section{Numerical Results and Discussion}
% \subsection{Simulation Configurations}
To evidence the performance of our proposed PPO-based CRS approach, we consider an underground pipeline monitoring scenario in Saudi Arabia for our simulations~\cite{LinMag}. Note that the proposed CRS-aided satellite-to-underground architecture can be generalized for other underground applications, such as smart agriculture and post-disaster rescue, by appropriately adjusting the channel model and system parameters. An LEO satellite equipped with $Q=6$ antenna elements serves $N=5$ single-antenna UDs, which are randomly distributed within a $1000$~m radius circular area and buried at a uniform depth of $d_u=0.6$~m. Meanwhile, a single-antenna AR with height $H_{ar} = 5$~m is located at the center of monitoring area to relay the received signals to the UDs. To ensure realistic modeling, the \textit{in-situ} clay percentage of the soil obtained from~\cite{SoilPara} is used to calculate the underground path loss. The carrier frequency is set to $433$~MHz, which is typical for underground wireless communication~\cite{LinBSWPUSNs}, and the LoRa modulation scheme is employed with a noise power of $-117$~dBm~\cite{LinMag}. The LEO-to-AR and LEO-to-UD channels are modeled as line-of-sight with a Rician factor and path loss exponent of $10$ and $2$, respectively, while the relay-to-UD channels follow non-line-of-sight propagation with a Rician factor and path loss exponent of $3$ and $2.4$~\cite{LinMag, Undergroundfield}. The specific simulation parameters are listed in Table~\ref{tab1}. In the proposed PPO framework, the actor and critic networks share a common feature extractor composed of two fully connected hidden layers with $512$ and $256$ neurons, respectively, each followed by layer normalization and GELU activation. The parameters of all neural networks are optimized using the AdamW optimizer~\cite{AdamW}. The other hyper-parameter settings in training process are summarized in Table~\ref{tab1}.

For performance comparison, three benchmark schemes are implemented:

\begin{itemize}
    \item \textbf{PPO-based SDMA.} The PPO-based SDMA approach in~\cite{PPOSDWCL} is extended to the considered satellite-to-underground scenario, where SDMA is employed for downlink transmission, and power allocation is optimized using the proposed PPO algorithm.

    \item \textbf{PPO-based RSMA.} Based on~\cite{PPOLEOWCL}, the PPO-based RSMA adopts a classical one-layer RSMA strategy for LEO-to-UD downlink communication without an AR, where the PPO algorithm jointly optimizes the power and rate allocations of the common and private streams.

    \item \textbf{Greedy-based CRS.} A greedy algorithm is employed to solve the CRS max–min rate optimization problem, where all historical rewards are stored, and the agent selects the action that yields the highest reward among past experiences~\cite{GreedyTcom}.
\end{itemize}

\begin{table}[t]
\caption{Simulation Parameters} 
\label{tab1}
\centering
\renewcommand{\arraystretch}{1} % 调整行高为默认的90%
\begin{tabular}{m{0.28\textwidth}<{\raggedright} m{0.15\textwidth}<{\centering}}%
\toprule
\textbf{Parameters}                & \textbf{Values}                    \\  \hline
\multicolumn{2}{l}{\textbf{Operation Environments}}\\ \hline
Radius of deployment area            & 1000~m               \\
Total number of UDs ($N$)                    & var (5 by default)                          \\
Burial depth ($d_u$)                 & var (0.6~m by default)                               \\
VWC ($m_v$)                          & var (15\% by default)                        \\
Clay ($m_c$)                         & 16.86\(\%\)                         \\ %\hline
% \multicolumn{2}{l}{\textbf{LEO Satellite Configuration}}       \\ \hlineOrbital height ($H_{S}$)             & 550~km                     \\
Antenna number of LEO satellite($Q$)             & 6     \\
Transmit power of LEO satellite ($P_t$)                    & 30~dBm                      \\
Antenna gain of LEO satellite ($G_s$)                    & 22.6~dBi                     \\
                        %\hline
% \multicolumn{2}{l}{\textbf{Relay Node Configuration}}        \\ \hline 
Height of AR ($H_{ar}$)               & 5~m                     \\
Transmit power of AR ($P_{R}$)              & 20~dBm                  \\
Antenna gain of AR  ($G_{ar}$)               & 5~dBi                    \\
Antenna gain of UDs ($G_n$)         &2.15~dBi                     \\%\hline
% \multicolumn{2}{l}{\textbf{Radio Configuration}} \\   \hline
Carrier frequency ($f$)  & 433~MHz                    \\
Noise power                  & -117~dBm                  \\
Rician factor LEO-to-UDs, LEO-to-AR, and AR-to-UDs channels              &10, 10, 3     \\
Path loss exponents for LEO-to-UDs, LEO-to-AR, and AR-to-UDs channels   &2, 2, 2.4     \\ \hline
\multicolumn{2}{l}{\textbf{PPO Configurations}} \\ \hline
Total epoch step ($T_{e}$)  &2000           \\
Batch size ($T_{b}$)  &512            \\
Update frequency of neural networks ($M$)  &3   \\
Learning rate ($\tau$)      &0.0001         \\
Discount factor ($\eta$)    &0.9            \\
GAE smoothing parameter ($\varsigma$) &0.95         \\
Clipping value ($\epsilon$)     &0.2            \\ 
\bottomrule
\end{tabular}
\renewcommand{\arraystretch}{1} % 恢复默认行高
\end{table}

% \subsection{Results Analysis}
\begin{figure}[!t]
\centering
\includegraphics[width=2.4in]{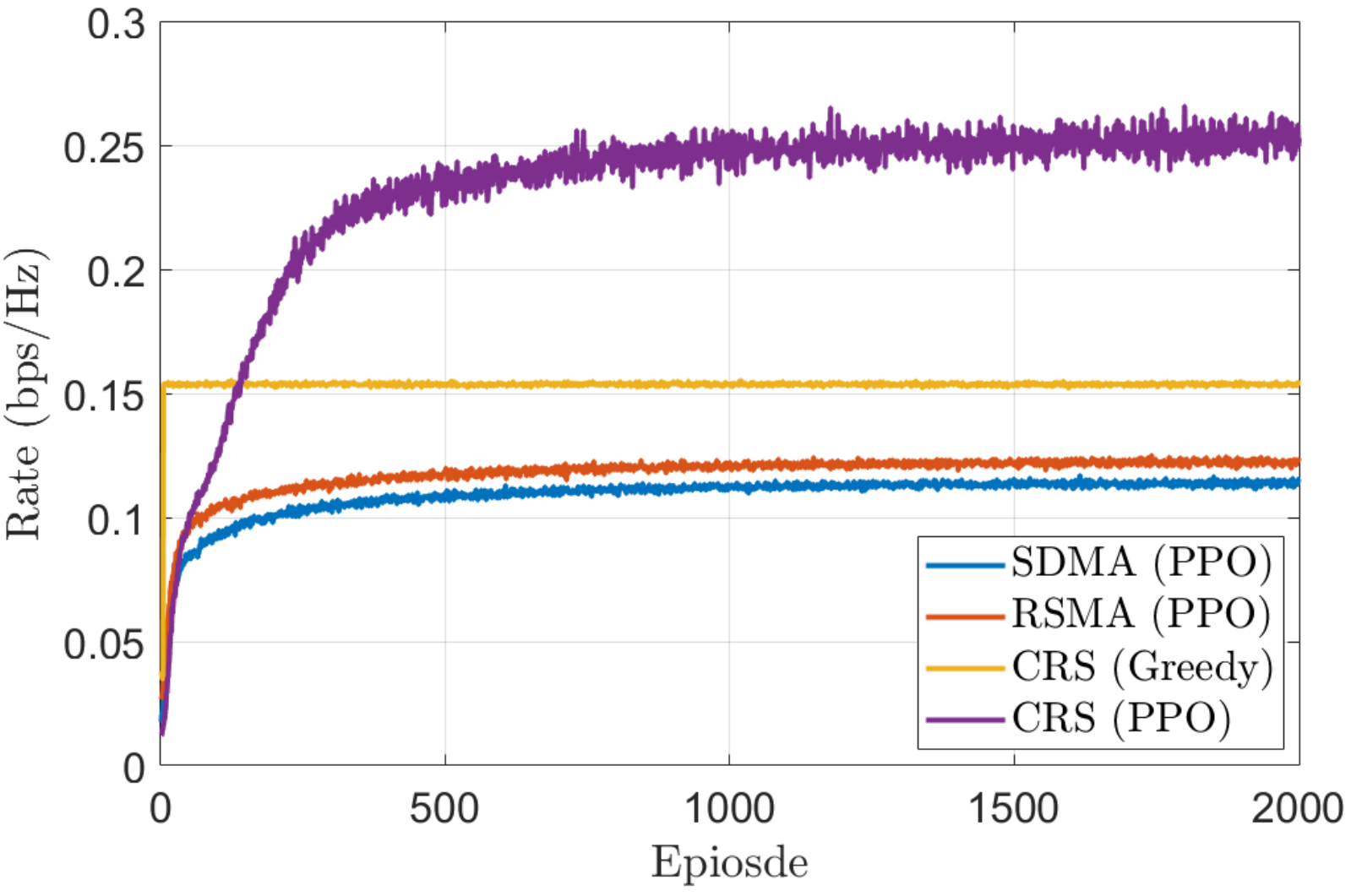}
\caption{Convergence performance of the proposed PPO algorithm for the SDMA, RSMA, and CRS strategies, as well as the greedy-based CRS scheme.}
\label{fig_2}
\vspace{-6mm}
\end{figure}

\begin{figure*}[!t]
\centering
\includegraphics[width=6.6in]{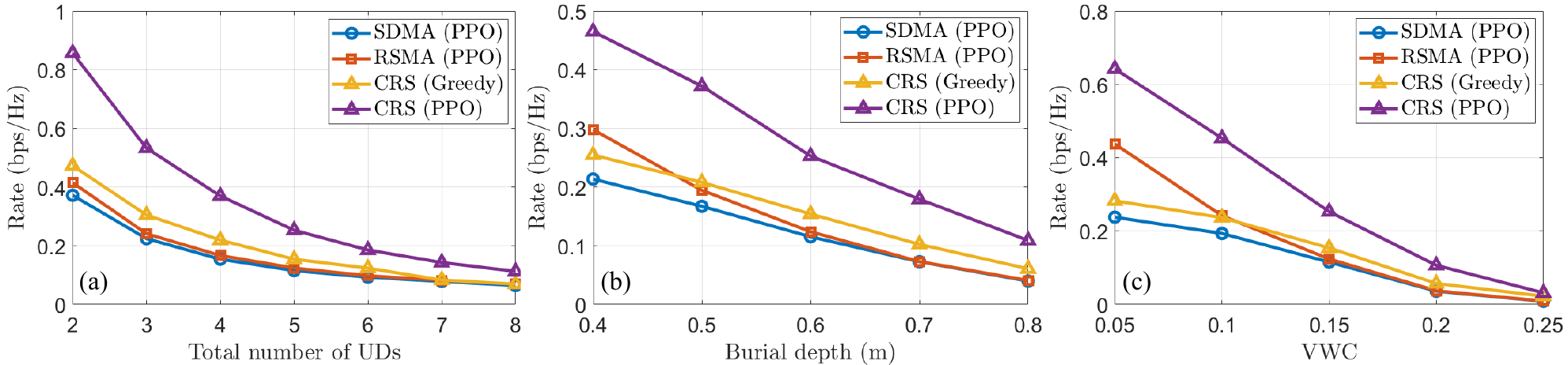}
\caption{Max–min rate performance versus (a) the number of UDs, (b) the burial depth of UDs, and (c) the soil's VWC for the PPO-based SDMA, PPO-based RSMA, greedy-based CRS, and PPO-based CRS strategies, averaged over $512$ random channel realizations.}
\label{fig_3}
\vspace{-4mm}
\end{figure*}

We first present the average reward results (i.e., the minimum rate among UDs) during the training process for SDMA, RSMA, and PPO-based CRS strategies in Fig.~\ref{fig_2}. One can observe that the PPO algorithm converges to a stable value within the first $500$ training episodes for both SDMA and RSMA strategies. In contrast, the CRS strategy requires nearly $1000$ episodes to converge due to its larger state and action space. This complexity arises from the joint optimization of power and rate allocation for the AR and all UDs, as well as the time slot allocation $\theta$. The greedy algorithm converges within only a few steps since it purely exploits historical rewards without exploration, whereas PPO requires more iterations to gradually balance exploration and exploitation for achieving a near-optimal policy. Upon convergence, the average reward achieved by the PPO-based CRS strategy is up to $219\%$, $204\%$, and $164\%$ higher than those of the PPO-based SDMA, PPO-based RSMA, and greedy-based CRS strategies, respectively. These results also demonstrate that the proposed DRL architecture can be effectively generalized to SDMA and RSMA strategies.

Fig.~\ref{fig_3} depicts the average max-min rate performance of different strategies under varying numbers of UDs, burial depths, and soil's VWC levels. Fig.~\ref{fig_3}(a) reveals that the worst-case rate decreases as the number of UDs increases due to increased competition for limited resources and a higher probability of UDs experiencing poor channel conditions. The proposed PPO-based CRS approach outperforms the benchmark schemes, achieving average performance gains of $212\%$, $197\%$, and $168\%$ over the PPO-based SDMA, PPO-based RSMA, and greedy-based CRS strategies, respectively, across all UDs' number scenarios. Fig.~\ref{fig_3}(b) illustrates that the minimum rate declines as the burial depth increases from $0.4$~m to $0.8$~m, since the longer propagation path through underground soil leads to heightened attenuation. The proposed PPO-based CRS achieves an average worst-case rate of $0.11$~bps/Hz at a burial depth of $0.8$~m, which is $274\%$, $267\%$, and $179\%$ higher than those of the PPO-based SDMA, PPO-based RSMA, and greedy-based CRS strategies, respectively. Fig.~\ref{fig_3}(c) shows that the average worst-case rate deteriorates with increasing VWC, as higher VWC results in a larger soil's attenuation constant, which significantly exacerbates underground signal attenuation and refraction loss in the air–soil interface. Nevertheless, the proposed PPO-based CRS approach consistently outperforms both benchmarks, thanks to its adaptive AR-assisted transmission and optimized resource allocation. For instance, at a VWC of $0.25$, the average worst-case rate achieved by our proposed approach improves by $371\%$, $347\%$, and $139\%$ compared to the PPO-based SDMA, PPO-based RSMA, and greedy-based CRS strategies, respectively. Furthermore, the performance gains of the CRS framework over the SDMA and RSMA schemes become more pronounced at greater burial depths and higher VWC levels, as the introduction of the AR and PPO-based resource optimization effectively mitigates the severe signal attenuation in soil and the refraction loss in the air–soil interface.

\section{Conclusion}
This paper proposed a CRS-aided satellite-to-underground communication system and employed a PPO algorithm to efficiently solve the max-min fairness problem that jointly optimizes power allocation, message splits, and time slot scheduling under uncertain channel conditions. Through comparisons with two benchmark schemes in a realistic underground pipeline monitoring case, our numerical results demonstrated that the proposed approach achieves superior max-min rate performance over three benchmarks. This work shows that the DRL-based CRS transmission framework is attractive to enable reliable satellite-to-underground communication.  

\bibliographystyle{IEEEtran}
\bibliography{ref}

\end{document}